\documentclass[12pt]{article}%
\usepackage{amsmath}
\usepackage{amsfonts}
\usepackage{amssymb}
\usepackage{graphicx}%
\begin{document}

\title{Anomalous energy losses in fractal medium}
\author{Sergey Panyukov, Andrei Leonidov\\\; \\Theoretical Physics Department \\P.N. Lebedev Physics Institute, Moscow, Russia}
\maketitle

\begin{abstract}
We derive equation describing distribution of energy losses of the particle
propagating in fractal medium with quenched and dynamic heterogeneities. We
show that in the case of the medium with fractal dimension $2<D<3$ the losses
of energy are described by the Mittag-Leffler renewal process. The average
energy loss of the particle experiences anomalous drift $\Delta\sim x^{\alpha
}$ with power-law dependence on the distance $x$ from the surface and exponent
$\alpha=D-2$.

\end{abstract}

\newpage

\section{Introduction}

In this paper we study the problem of energy losses of the particle
propagating in disordered medium. We consider both a model of discrete energy
loss, when the particle collides with randomly placed scattering centers in
the medium, so that during intervals between such scattering events the
particle has ballistic trajectory and the energy of the particle does not
change, and its continuum limit, corresponding to the description of the
energy loss process in terms of kinetic equation.

There exists a close analogy between this problem and the diffusion of a
particle through the medium: the energy loss is proportional to the number of
scattering events $n\left(  x\right)  $ at the distance $x$ from the surface
of the medium, and $x$ plays the role of the time $t$ in the diffusion
problem. The propagation of a particle in the regular homogeneous medium, with
the number of scattering events $\left\langle n\right\rangle \sim x$, is
analogous to the standard Brownian motion with the average number of events
$\left\langle n\right\rangle \sim t$. Anomalous power-law coordinate
dependence $\left\langle n\right\rangle \sim x^{\alpha}$ is related to the
presence of power-law spatial correlations in the density distribution of
scattering centers in the random medium. Such correlations are typical both
for fractal systems with quenched disorder and for systems with dynamic
heterogeneities at the critical point of a phase transition.

In diffusion problems anomalous time dependence of the number of events is
often observed near glass transition\cite{Glass1,Glass2}, for
sub-diffusion-limited reactions\cite{Reactions}, in turbulent
regime\cite{Turbulence} and for some transport problems in disordered
systems.\cite{Disorder1,Disorder2}

Our consideration of the discrete energy loss model with power law
correlations is done at the level analogous to the Continuous Random Walk
(CTRW) \cite{CTRW_model}. The main distinction of the stochastic process
describing the energy loss and the CTRW one is that the energy loss problem
corresponds to a sum of random positive quantities -- energy losses at
scattering events. Of substantial interest is a continuous limit of the
discrete energy loss model corresponding to some effective kinetic
description. The highly nontrivial and interesting transition from the
discrete to kinetic description in the usual CTRW case was discussed in
\cite{H95}.

In section~\ref{CONTINUOUS} we study the energy losses in weakly heterogeneous
medium. We demonstrate that average energy losses in such medium can
anomalously grow with the distance $x$ from the surface, $\left\langle
\Delta(x)\right\rangle \sim x^{1/\beta}$, with \emph{super}-diffusional
exponent $\beta^{-1}>1$. In section~\ref{INHOMO} we study energy losses in
strongly heterogeneous medium. We first formulate simple model of anomalous
scattering, describing the loss of energy by the particle during its
propagation in random medium (section~\ref{CONSTANT}). In section~\ref{CTRW}
this process is generalized to the case of general distribution of energy
losses during individual collisions using formalism of CTRW. We derive
equation for the distribution function of energy losses and show that the
energy of the particle experiences anomalous drift with average energy losses
growing with the distance $x$ as $\left\langle \Delta(x)\right\rangle \sim
x^{\alpha}$, with \emph{sub}-diffusional exponent $\alpha<1$. In
section~\ref{CONT} we demonstrate, that the loss of the energy because of
scattering on fractal structure is described by the Mittag-Leffler renewal
process\cite{Renewal}, which may be considered as fractional generalization of
the well known Poisson renewal process. Main results of this work are
summarized in section~\ref{Conclusion}.

\section{Weakly heterogeneous medium\label{CONTINUOUS}}

The loss of the energy by the particle propagating in homogeneous medium is
described, in the continuous limit, by the Landau kinetic equation for the
distribution function of energy losses at point $x$\cite{Book}%
\begin{equation}
\frac{\partial f\left(  \Delta\mathbf{,}x\right)  }{\partial x}=\frac{1}{a}%
{\displaystyle\int_{0}^{\infty}}
d\varepsilon w\left(  \varepsilon\right)  \left[  f\left(  \Delta
-\varepsilon,x\right)  -f\left(  \Delta,x\right)  \right]  \label{bk0}%
\end{equation}
where $w\left(  \varepsilon\right)  $ is the probability distribution of
energy loss at scattering event and $a^{-1}$ is linear density of scattering
centers separated by a distance $a$.

In the random medium the probability distribution of energy loss becomes a
random function of the coordinate $x$, $w\left(  \varepsilon\right)
\rightarrow w\left(  \varepsilon|x\right)  $. In principle one could imagine
arbitrary variations of the form of $w\left(  \varepsilon|x\right)  $ from one
point to another that could be correlated over spatial domains with the size
controlled by the corresponding correlation length $\xi$.

In the simplest case of finite correlation length $\xi$ the basic kinetic
equation for the averaged distribution function $f\left(  \Delta
\mathbf{,}x\right)  $ on distances $x\gg\xi$ takes the form
\begin{equation}
\frac{\partial f\left(  \Delta\mathbf{,}x\right)  }{\partial x}=\frac{1}{a}%
{\displaystyle\int_{0}^{\infty}}
d\varepsilon{\bar{w}}(\varepsilon)\left[  f\left(  \Delta-\varepsilon
,x\right)  -f\left(  \Delta,x\right)  \right]  \label{bk}%
\end{equation}
with
\begin{equation}
{\bar{w}}(\varepsilon)=\langle w(\varepsilon|x)\rangle. \label{avdisfun}%
\end{equation}
The averaging in Eq. (\ref{avdisfun}) is over the randomness of the
distribution function $w(\varepsilon|x)$.

As an illustration, let us consider the distribution $w(\varepsilon|\sigma)$
used in Eq.~(\ref{bk0}), where $\sigma$ denotes the set of parameters
characterizing this distribution. The simplest way of introducing spatial
randomness of the distribution $w(\varepsilon|\sigma)$ is to consider the
parameters $\sigma$ as random fields depending on spatial coordinates,
$w(\varepsilon|\sigma)\rightarrow w(\varepsilon|\sigma(x))$. At distances
larger than the correlation length $\xi$ the fields $w(\varepsilon|\sigma(x))$
are effectively uncorrelated and the distribution ${\bar{w}}(\varepsilon)$ is
simply given by
\begin{equation}
{\bar{w}}(\varepsilon)=\int d\sigma p(\sigma)w(\varepsilon|\sigma) \label{dd}%
\end{equation}
where $p(\sigma)$ is the probability distribution describing the local
fluctuations of $\sigma(x)$. The properties of the resulting distribution
${\bar{w}}(\varepsilon)$ can substantially differ from those of the initial
distribution $w(\varepsilon|\sigma)$. Of special importance is the influence
of randomness in $\sigma$ on the asymptotic behavior of ${\bar{w}}%
(\varepsilon)$ at large $\varepsilon$. In particular, the power-like decay of
$p(\sigma)$ at large $\sigma$ induces the power-like decay of ${\bar{w}%
}(\varepsilon)$ even for well-localized exponentially decaying $w(\varepsilon
|\sigma)$.

The solution of Eq.~(\ref{bk}) is most conveniently obtained by solving for
the Laplace-transformed distribution function
\begin{equation}
{\tilde{f}}(p,x)\equiv\int_{0}^{\infty}d\Delta e^{-p\Delta}f(\Delta
,x)=\exp\left[  -\omega(p)x\right]  \label{Expon}%
\end{equation}
where
\begin{equation}
\omega\left(  p\right)  =\frac{1}{a}\int_{0}^{\infty}d\varepsilon{\bar{w}%
}\left(  \varepsilon\right)  \left(  1-e^{-p\varepsilon}\right)  \label{omega}%
\end{equation}

Calculating the inverse Laplace transform of Eq.~(\ref{Expon}) we find that
the distribution function $f\left(  \Delta\mathbf{,}x\right)  $ can have
different form depending on the large energy asymptotes of the distribution
${\bar{w}}\left(  \varepsilon\right)  $:

a) If this function ${\bar{w}}\left(  \varepsilon\right)  $ decays at large
$\varepsilon$ faster than $1/\varepsilon^{3}$, the distribution function
$f\left(  \Delta\mathbf{,}x\right)  $ has Gaussian form with the center at
$\Delta=\Delta_{1}x$ and the width $\sqrt{\Delta_{2}x}$, where $\Delta_{k}$
are corresponding moments of the distribution ${\bar{w}}\left(  \varepsilon
\right)  $:%
\begin{equation}
\Delta_{k}=\int_{0}^{\infty}d\varepsilon\varepsilon^{k}{\bar{w}}\left(
\varepsilon\right)  \label{d0}%
\end{equation}

b) If the function $\bar{w}\left(  \varepsilon\right)  $ decays as
$1/\varepsilon^{1+\beta}$ with $1<\beta<2$ the distribution function $f\left(
\Delta\mathbf{,}x\right)  $ is still centered at $\Delta=\Delta_{1}x$ but is
not Gaussian, and its width grows as power law $x^{1/\beta}$ of $x$. In both
cases a) and b) the distribution $f\left(  \Delta\mathbf{,}x\right)  $ becomes
sharper with the rise of $x$.

c) If the function ${\bar{w}}\left(  \varepsilon\right)  $ decays slower than
$1/\varepsilon^{2}$ then the first moment $\Delta_{1}$ diverges. In the case
of power distribution ${\bar{w}}\left(  \varepsilon\right)  \sim$
$\varepsilon^{-1-\beta}$ with exponent $0<\beta<1$ we find from
Eq.~(\ref{omega}) $\omega\left(  p\right)  =Cp^{\beta}$ with certain constant
$C\sim\Delta_{1}^{\beta}/a$. Calculating the inverse Laplace transformation of
Eq.~(\ref{Expon}) we get%
\begin{equation}
f\left(  \Delta\mathbf{,}x\right)  =\beta\frac{Cx}{\Delta^{1+\beta}}W_{\beta
}\left(  \frac{Cx}{\Delta^{\beta}}\right)  , \label{fb}%
\end{equation}
where $W_{\beta}\left(  z\right)  $ is the Wright type function (see
Appendix~\ref{WRIZE}). The center of this distribution and its width are on
the same order of value and grow as the power law $\left\langle \Delta\left(
x\right)  \right\rangle \sim x^{1/\beta}$ of the distance $x$ from the
surface. Such dependence is familiar for L\'{e}vy flight processes
characterized by the exponent $1/\beta>1$.\cite{Levy}

\section{Strongly heterogeneous medium\label{INHOMO}}

In this section we consider energy losses in strongly heterogeneous medium
with infinite correlation radius $\xi$ (or on distances from the surface small
with respect to $\xi$). Such situation takes place in the case of fractal
medium or for the system at critical point of phase transition.

\subsection{Simple model with constant losses\label{CONSTANT}}

\begin{figure}
[tbh]
\begin{center}
\includegraphics[
height=0.4367in,
width=2.3354in
]%
{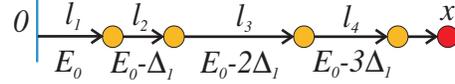}%
\caption{Particle propagating in random medium looses the energy $\Delta_{1}$
during scattering events which are separated by random intervals $l$ along the
trajectory of the particle.}%
\label{toy1}%
\end{center}
\end{figure}

Let us first consider the simple one-dimensional microscopic model for energy
loss in heterogeneous medium in which the loss takes place at through a
sequence of discrete events in which the incident particle looses an equal
amount of energy $\Delta_{1}$. This model describes the eikonal energy loss of
a high energy particle loosing energy in small portions so that one can, in
the first approximation, consider its trajectory to be a straight line. The
model generalizes the simplest model of discrete random walk with the fixed
elementary step to the case of the positive variable. The assumption of
constant energy loss will be relaxed in the next paragraph in which we will
consider a more general model. More precisely, in the simple model we consider
a particle entering the medium at the point $x=0$ and loosing energy at random
points $x_{1},\cdots,x_{n}$ lying on the trajectory composed by the intervals
of length $l_{1},\cdots,l_{n}$, see Fig. \ref{toy1}.

The cumulative energy loss at some point $x$ for the given event is,
evidently, equal $\Delta\left(  x\right)  =\Delta_{1}n\left(  x\right)  $,
where $n(x)$ is the number of scattering events that the particle experienced
along its trajectory from the initial point to the point under consideration.
The probabilistic description of the process of random energy loss is thus
fully specified by the probabilistic properties of $n(x)$. Let consider the
case in which these properties are fully described by the probability density
$\psi(l)$ of the spatial distance $l\equiv x-x^{\prime}$ between the points
$x$ and $x^{\prime}<x$ at which two subsequent scattering events took place
and the distribution $\varphi(l_{1})$ characterizes the probability density of
the position $l_{1}$ of the first scattering event, see Fig.~\ref{toy1}. In
what follows we will simply assume that $\varphi(l_{1})=\psi(l_{1})$. The
function $\psi(l)$ determines the survival probability $\Psi\left(  r\right)
$ that the particle experiences no collisions at the distance $r=x-x^{\prime}$
after the collision at point $x^{\prime}$:%
\begin{equation}
\Psi\left(  r\right)  =\int_{r}^{\infty}\psi(l)dl \label{Psi}%
\end{equation}

To develop a quantitative description of the energy loss process it is
convenient to introduce the probability density $\psi_{n}\left(  x-x^{\prime
}\right)  $ for the $n$-th collision to take place at the distance
$x-x^{\prime}$ from the collision at point $x^{\prime}$. Then the probability
density $f_{n}(x)$ for the number $n$ of scattering events along the
trajectory leading to the point $x$ reads
\begin{equation}
f_{n}\left(  x\right)  =\int_{0}^{x}dx^{\prime}\Psi\left(  x-x^{\prime
}\right)  \psi_{n}\left(  x^{\prime}\right)  \label{PN}%
\end{equation}
In turn, the probability density $\psi_{n}\left(  x\right)  $ is determined by
recurrence relation%
\begin{equation}
\psi_{n}\left(  x\right)  =\int_{0}^{x}dx^{\prime}\psi\left(  x-x^{\prime
}\right)  \psi_{n-1}\left(  x^{\prime}\right)  \label{psik}%
\end{equation}
with $\psi_{1}\left(  x\right)  =\psi\left(  x\right)  $.

The equations (\ref{PN},\ref{psik}) are easily solved by using the Laplace
transform
\[
\tilde{f}_{n}\left(  q\right)  =\int_{0}^{\infty}f_{n}\left(  x\right)
e^{-qx}dx
\]
We have
\begin{equation}
\tilde{f}_{n}\left(  q\right)  =\tilde{\psi}^{n}\left(  q\right)
\frac{1-\tilde{\psi}\left(  q\right)  }{q} \label{fn}%
\end{equation}
where $\tilde{\psi}\left(  q\right)  $ is the Laplace transform of the
probability density $\psi(l)$.

Of special interest is the case of the medium characterized by power-law
scale-invariant fluctuations of the positions of the scattering centers. An
example of this situation is provided by the fractal medium. In the considered
one-dimensional case the relevant correlations are determined by projecting
the full three-dimensional correlation pattern on the axis of particle
propagation. In the scale-invariant case the function $\psi(l)$ should read
\begin{equation}
\psi\left(  l\right)  =\left\{
\begin{array}
[c]{ccc}%
0 & \text{at} & l<a\\
\alpha a^{\alpha}/l^{1+\alpha} & \text{at} & l>a
\end{array}
\right.  . \label{psiA}%
\end{equation}
where, as shown in Appendix~\ref{EXPONENT}, exponent $\alpha$ is related to
the fractal dimension $D$ of the scattering medium as:%
\begin{equation}
\alpha=D-2.
\end{equation}
Notice, that there are both upper and low boundaries of the fractal dimension
$D$ for the problem of anomalous scattering. Upper fractal dimension $D=3$
describes homogeneous medium and at lower critical fractional dimension $D=2$
the scattering cluster is too sparse to effectively scatter the propagating particle.

At distances much larger than the ultraviolet cutoff $a$ the Laplace transform
of the function in Eq.~(\ref{psiA}) reads
\begin{equation}
\tilde{\psi}\left(  q\right)  =\int_{0}^{\infty}\psi\left(  x\right)
e^{-qx}dx\simeq1-\left(  aq\right)  ^{\alpha},\quad aq\ll1 \label{psi_q}%
\end{equation}
In the case $\alpha<1$ the average scattering length $\bar{l}$ diverges
leading to anomalous losses of the particle energy. Calculating the inverse
Laplace transform of Eq.~(\ref{fn}), we get at large $x\gg a$ the distribution
function of energy losses:
\begin{equation}
f_{n}\left(  x\right)  =\left(  a/x\right)  ^{\alpha}W_{\alpha}\left[  \left(
a/x\right)  ^{\alpha}n\right]  , \label{fnx}%
\end{equation}
where $W_{\alpha}\left(  z\right)  $ is the Wright type function, defined in
Eq.~(\ref{Wa}) of Appendix~\ref{WRIZE}. Using this distribution, we can
calculate the average energy loss at the distance $x$ from the surface, which
is monotonically increasing function of $x$:%
\begin{equation}
\left\langle \Delta\left(  x\right)  \right\rangle =\Delta_{1}\bar{n}\left(
x\right)  =\frac{\Delta_{1}}{\Gamma\left(  1+\alpha\right)  }\left(  \frac
{x}{a}\right)  ^{\alpha},\qquad0<\alpha<1 \label{Dav}%
\end{equation}
At $\alpha=1$ the average scattering length $\bar{l}$ is finite and in average
the losses of particle energy $\left\langle \Delta(x)\right\rangle =\Delta
_{1}x/\bar{l}$ grow linearly with the distance $x$, similar to the case of
homogeneous system \cite{Book}.

\subsection{Continuous Time Random Walk (CTRW)\label{CTRW}}

In this section we consider the case, when energy loss $\varepsilon$ at each
collision with scattering center is random and is described by the
distribution function $w\left(  \varepsilon\right)  $. By analogy with CTRW
\cite{CTRW_model} the process of energy losses can be considered as
subordinated to the anomalous scattering process shown in Fig.~\ref{toy1}. Let
us stress once again that the model under consideration differs from the usual
CTRW in considering the random walk of the positive variable $\varepsilon$.
The distribution function of energy losses can be presented as the convolution
of the distribution function of the number of collisions $f_{n}\left(
x\right)  $ studied in previous section and the distribution function of
energy loss $w_{n}\left(  \Delta\right)  $ for the given number of collisions
$n$:
\begin{equation}
f\left(  \Delta\mathbf{,}x\right)  =\sum_{n}w_{n}\left(  \Delta\right)
f_{n}\left(  x\right)  . \label{Convol}%
\end{equation}
Due to the additive nature of the energy loss the distribution function
$w_{n}\left(  \Delta\right)  $ is determined by the inverse Laplace transform
of the power $\tilde{w}^{n}\left(  p\right)  $ of the Laplace transform
$\tilde{w}\left(  p\right)  $ of the distribution $w\left(  \varepsilon
\right)  $. Explicit expression~(\ref{Convol}) for the distribution function
$f\left(  \Delta\mathbf{,}x\right)  $ is not very convenient to use, so below
we derive the kinetic equation for it.

Simple probabilistic consideration of the losses of the energy by the particle
propagating in random medium characterized by the probability distribution
$w\left(  \varepsilon\right)  $ shows that this random process can be
described by equation%
\begin{equation}
f\left(  \Delta\mathbf{,}x\right)  =\delta\left(  \Delta\right)  \Psi\left(
x\right)  +\int_{0}^{x}dx^{\prime}\psi\left(  x-x^{\prime}\right)  \int
_{0}^{\infty}d\varepsilon w\left(  \varepsilon\right)  f\left(  \Delta
-\varepsilon\mathbf{,}x^{\prime}\right)  \label{fDx}%
\end{equation}
where $\Psi\left(  x\right)  $ is the survival probability at the distance $x$
from the surface, Eq.~(\ref{Psi}). To solve Eq.~(\ref{fDx}) it is convenient
to introduce the double Laplace transform of $f\left(  \Delta\mathbf{,}%
x\right)  $ over both variables $\Delta$ and $x$%
\[
\tilde{f}\left(  p,q\right)  \equiv\int_{0}^{\infty}d\Delta e^{-p\Delta}%
\int_{0}^{\infty}dxe^{-qx}f\left(  \Delta,x\right)
\]
The corresponding equation for $\tilde{f}\left(  p,q\right)  $ reads:%
\begin{equation}
\tilde{f}\left(  p,q\right)  =\frac{1-\tilde{\psi}\left(  q\right)  }%
{q}+\tilde{\psi}\left(  q\right)  \tilde{w}\left(  p\right)  \tilde{f}\left(
p,q\right)
\end{equation}
Introducing the function%
\begin{equation}
\tilde{g}\left(  q\right)  =\frac{\tilde{\psi}\left(  q\right)  }%
{1-\tilde{\psi}\left(  q\right)  } \label{g_q}%
\end{equation}
we can rewrite the above equation (\ref{g_q}) in the form%
\begin{equation}
\tilde{f}\left(  p,q\right)  =\frac{1}{q}+\tilde{g}\left(  q\right)  \left[
\tilde{w}\left(  p\right)  -1\right]  \tilde{f}\left(  p,q\right)  \label{ff0}%
\end{equation}
which is equivalent to%
\begin{align}
f\left(  \Delta\mathbf{,}x\right)   &  =\delta\left(  \Delta\right)  +%
{\displaystyle\int_{0}^{x}}
dx^{\prime}g\left(  x-x^{\prime}\right)  \times\label{ff1}\\
&
{\displaystyle\int_{0}^{\infty}}
d\varepsilon w\left(  \varepsilon\right)  \left[  f\left(  \Delta
-\varepsilon,x^{\prime}\right)  -f\left(  \Delta,x^{\prime}\right)  \right]
\nonumber
\end{align}
Eq.~(\ref{ff1}) can be considered as generalization of Eq.~(\ref{bk0}) for
homogeneous medium corresponding to constant density of scattering centers,
$g\left(  r\right)  =1/a$. The function\textbf{\ }$g\left(  r\right)
$\textbf{\ }is found by inverse Laplace transform of the function $\tilde
{g}\left(  q\right)  $, Eq.~(\ref{g_q}). Using Eq.~(\ref{fn}) $\tilde
{g}\left(  q\right)  $ can be related to the Laplace transform of the average
number of scattering events:%
\begin{equation}
\tilde{n}\left(  q\right)  =\sum\nolimits_{n}n\tilde{f}_{n}\left(  q\right)
=\frac{\tilde{g}\left(  q\right)  }{q},
\end{equation}
We conclude, that $g\left(  r\right)  $ has the meaning of average density of
scattering events along the direction $\mathbf{e_{x}}$ of particle propagation
at the distance $r$ from the scattering event:%
\begin{equation}
g\left(  r\right)  =d\bar{n}\left(  r\right)  /dr \label{gxn}%
\end{equation}

In general, the function $g\left(  r\right)  $ depends on characteristics of
the medium, and it can be related
\begin{equation}
g\left(  r\right)  =a^{2}G\left(  \mathbf{e_{x}}r\right)  , \label{gG}%
\end{equation}
to the so-called structure function of the medium
\begin{equation}
G\left(  \mathbf{r}\right)  =\left\langle \sum\nolimits_{n\neq0}\delta\left(
\mathbf{x}_{i}-\mathbf{x}_{i+n}-\mathbf{r}\right)  \right\rangle , \label{SF}%
\end{equation}
$\mathbf{x}_{i}$ are coordinates of the $i$-th scattering center. In
Eq.~(\ref{gG}) $a^{2}$ is the scattering area of the particle and
$\mathbf{e_{x}}$ is unit vector along the trajectory of the particle. This
relation can be established rewriting Eq.~(\ref{SF}) in the form
\begin{align}
G\left(  \mathbf{r}\right)   &  =\sum_{n\neq0}\psi_{n}^{\prime}\left(
\mathbf{r}\right)  ,\quad\int\psi_{n}^{\prime}\left(  \mathbf{r}\right)
d\mathbf{r}=1\label{Struct}\\
\psi_{n}^{\prime}\left(  \mathbf{r}\right)   &  =\left\langle \delta\left(
\mathbf{x}_{i}-\mathbf{x}_{i+n}-\mathbf{r}\right)  \right\rangle \nonumber
\end{align}
Although the sum in Eq.~(\ref{Struct}) is going over all scattering centers in
the medium, only centers along the trajectory of the particle enter into
Eq.~(\ref{gG}). Expanding Eq.~(\ref{g_q}) in powers of $\tilde{\psi}\left(
q\right)  $ and taking the inverse Laplace transform of each term of the
obtained series, we get%
\begin{equation}
g\left(  r\right)  =\sum_{n\neq0}\psi_{n}\left(  x\right)  ,\quad\int\psi
_{n}\left(  r\right)  dr=1 \label{Gn}%
\end{equation}
where $\psi_{n}\left(  r\right)  =\left\langle \delta\left(  x_{i}%
-x_{i+n}-r\right)  \right\rangle $ is the probability distribution of the
distance $r=x_{i+n}-x_{i}$ between $n$ consequent collisions along the
trajectory of the propagating particle (see Eq.~(\ref{psik})). Comparing
Eq.~(\ref{Struct}) and~(\ref{Gn}) term by term, we reproduce
relation~(\ref{gG}) between functions $g\left(  r\right)  $ and $G\left(
\mathbf{r}\right)  $.

\subsection{Continuous limit\label{CONT}}

In the case of scattering medium with fractal dimension $D$ the Laplace
transform of the function $g\left(  r\right)  $ in the long wavelenth limit
$aq\ll1$ has the form:%
\begin{equation}
\tilde{g}\left(  q\right)  =\left(  aq\right)  ^{-\alpha},\quad\alpha=D-2
\label{gq}%
\end{equation}
In Appendix~\ref{ASYMPT} we derive asymptotic solutions of Eq.~(\ref{ff1})
close to the surface and far from it. Close enough to the surface the
distribution of energies will have large peak at $\Delta=0$, describing
non-scattered particle%
\begin{equation}
f\left(  \Delta,x\right)  \simeq\delta\left(  \Delta\right)  \Psi\left(
x\right)  , \label{fPsi}%
\end{equation}
where the function $\Psi\left(  x\right)  $ is the survival probability that
the particle does not scatter at the depths smaller than $x$. Using large $p$
asymptotes of the function $\omega\left(  p\right)  \simeq1/a$ defined in
Eq.~(\ref{omega}), we find from Eq.~(\ref{fEa}) of Appendix~\ref{ASYMPT}:
\begin{equation}
\Psi\left(  x\right)  =E_{\alpha}\left[  \left(  x/a\right)  ^{\alpha}\right]
\label{Psix}%
\end{equation}
where the function $E_{\alpha}$ is defined in Eq.~(\ref{Ea}) of
Appendix~\ref{WRIZE}.

The process with survival probability $\Psi\left(  r\right)  =E_{\alpha
}\left(  r\right)  $ is known as the Mittag-Leffler renewal process, that is
described by fractional differential equation%
\begin{equation}
\frac{\partial^{\alpha}\Psi\left(  r\right)  }{\partial r^{\alpha}}%
=-\Psi\left(  r\right)  \label{ML}%
\end{equation}
where $\partial^{\alpha}/\partial x^{\alpha}$ is the Caputo fractional
derivative%
\begin{equation}
\frac{\partial^{\alpha}\Psi\left(  r\right)  }{\partial r^{\alpha}}=\frac
{1}{\Gamma\left(  1-\alpha\right)  }\int_{0}^{r}\frac{\Psi^{\prime}\left(
x\right)  }{\left(  r-x\right)  ^{\alpha}}dx,\quad0<\alpha<1. \label{Caputo}%
\end{equation}
It was introduced by Caputo in the later 1960s for modeling the energy
dissipation in the rheology of the Earth\cite{Caputo}.

Far enough from the surface the loss of the energy is determined by multiple
scattering processes, when we can expand
\[
\omega\left(  p\right)  \simeq p\frac{\Delta_{1}}{a},\quad\Delta_{1}=\int
_{0}^{\infty}d\varepsilon\varepsilon w\left(  \varepsilon\right)
\]
Calculating the inverse Laplace transform of Eq.~(\ref{fS}) in
Appendix~\ref{ASYMPT}, we get
\begin{equation}
f\left(  \Delta,x\right)  \simeq\frac{1}{\Delta_{1}}\left(  \frac{a}%
{x}\right)  ^{\alpha}W_{\alpha}\left[  \frac{\Delta}{\Delta_{1}}\left(
\frac{a}{x}\right)  ^{\alpha}\right]  \label{fdx}%
\end{equation}
where the Wrize function $W_{\alpha}$ is defined in Eq.~(\ref{Wa}) of
Appendix~\ref{WRIZE}.

The function $f\left(  \Delta,x\right)  $~(\ref{fdx}) is the solution of the
space-fractional drift equation of the order $\alpha$:%
\begin{equation}
\frac{\partial f\left(  \Delta,x\right)  }{\partial\Delta}=-\frac{a^{\alpha}%
}{\Delta_{1}}\frac{\partial^{\alpha}f\left(  \Delta,x\right)  }{\partial
x^{\alpha}}%
\end{equation}
This distribution function~(\ref{fdx}) describes pure renewal process with
anomalous exponent $\alpha$.\cite{Renewal} This process can be modelled as the
series of jumps of the energy $\Delta$ with the amplitude $\Delta_{1}$ each
happened at renewal points $x$ separated by random discrete intervals $l$
distributed according to the power law: $\psi\left(  l\right)  \sim
l^{-1-\alpha}$, see Fig.~\ref{toy1}.

In the continuous limit we can generalize the simple model of
section~\ref{CONSTANT} using the thinning procedure\cite{thinning}. In this
procedure for each positive $n$ a decision is made: the scattering event is
maintained with probability $p$ or it is deleted with probability $1-p$. In
the limit $p\rightarrow0$ the amplitude of scattering become smaller and
smaller, their number in a given span of space larger and larger, and the
ballistic trajectories between scattering events smaller and smaller. In this
limit there are no ballistic trajectories anymore and we come to continuous medium.

When the energy distribution $w\left(  \varepsilon\right)  $ decays quicker
than $\varepsilon^{-2}$ energy losses for the thinning procedure model are
described by asymptotically universal Mittag--Leffler distribution. This
distribution is characterized by the spatial scale $b\gg a$ at which the
particle is scattering with the probability about $1/2$. In the random medium
with power-low correlations the parameter $b$ determines the amplitude of the
Laplace transform of the function $g\left(  r\right)  $:%
\begin{equation}
\tilde{g}\left(  q\right)  =\left(  bq\right)  ^{-\alpha},\quad aq\ll1
\label{gq1}%
\end{equation}

The Mittag--Leffler asymptotic distribution corresponds to the model of
constant energy losses $\Delta=\Delta_{1}n$ (section~\ref{CONSTANT}) with
asymptotic function $\tilde{g}\left(  q\right)  $~(\ref{gq1}). Calculating the
inverse Laplace transform of expression~(\ref{fnq}) in Appendix~\ref{ASYMPT},
and renormalizing the spacial scale $a\rightarrow b$ we get:
\begin{equation}
f_{n}\left(  x\right)  =\frac{\left(  x/b\right)  ^{\alpha n}}{n!}E_{\alpha
}^{\left(  n\right)  }\left[  \left(  \frac{x}{b}\right)  ^{\alpha}\right]  .
\label{fnE}%
\end{equation}
where $E_{\alpha}^{\left(  n\right)  }$ is $n$-th derivative of the function
$E_{\alpha}$, Eq.~(\ref{Ea}) of Appendix~\ref{WRIZE}, and the average over
this distribution $\bar{n}\left(  x\right)  =\left(  x/b\right)  ^{\alpha
}/\Gamma\left(  1+\alpha\right)  $~(\ref{Dav}) determines average losses at
distance $x$ from the surface, $\left\langle \Delta\left(  x\right)
\right\rangle =\Delta_{1}\bar{n}\left(  x\right)  $. At $n=0$ the
Mittag--Leffler distribution~turns to the survival probability,
Eq.~(\ref{fPsi}), while in the limit $x\gg b$ it turns to the renormalized
function~(\ref{fdx}). In the case $\alpha=1$ the distribution~(\ref{fnE})
takes the well known Poisson form%
\[
f_{n}\left(  x\right)  =\frac{\bar{n}^{n}}{n!}e^{-\bar{n}},\qquad\bar{n}%
=\frac{x}{b}%
\]
and may be considered as generalization of Poisson distribution of scattering
events for the case of fractal medium.

In the case of slowly decaying energy distribution at scattering events
$w\left(  \varepsilon\right)  \sim$ $\varepsilon^{-1-\beta}$ ($0<\beta<1$) the
scattering on fractal structures has the form of convolution of L\'{e}vy
flight processes characterized by the exponent $\beta$ with anomalous scaling
processes characterized by the exponent $\alpha$. Such combined process is
described by fractional space-energy differential equation, and average energy
losses grow with the distance $x$ from the surface as $\left\langle
\Delta\left(  x\right)  \right\rangle \sim x^{\alpha/\beta}$. The effective
exponent $\alpha/\beta$ of this process can be smaller or larger than $1$,
depending on relation between exponents $\alpha$ and $\beta$.

\section{Conclusion\label{Conclusion}}

We study the loss of energy of the particle moving in fractal media and in the
system with dynamic heterogeneities formed at the critical point of phase
transition. We show that when the distribution of energy loss during
collisions $w\left(  \varepsilon\right)  $ quickly decays with the energy
$\varepsilon$ the distribution function of particle energies is universal and
depends only on fractal dimension $D$ of the medium. In the case $D=3$ spacial
heterogeneities only weakly affect the scattering process, which can be
described by the classical theory.\cite{Book} In the case $2<D<3$ spacial
heterogeneities change the character of the scattering, which can be described
by fractional differential equations of order $\alpha=D-2$. Nonlocal character
of fractional derivatives (see Eq.~(\ref{Caputo})) reflects power-law
correlations existing in the fractal system. We show that the loss of the
energy in fractal medium can be described by the Mittag-Leffler renewal
process of order $\alpha$, which is fractional generalization of the Poisson
process corresponding to the case $\alpha=1$ of absence of such correlations.

One of the most important applications of this theory is propagation of
particle through the percolation scattering structure with fractal dimension
$D\simeq2.49$, when the exponent $\alpha\simeq0.49$. Similar exponent
$\alpha=1/2$ is obtained for fractals with lattice animals structure -- a set
of randomly connected sites on a lattice.\cite{Animal} The lower boundary
$D=2$ (corresponding to random walk structures) of applicability of our
consideration equal to the dimension $d=2$ of the interface. Therefore in the
case of fractal dimension $D<2$ of scattering clusters the losses of the
particle energy per unit area decrease with the rise of the interface area.

We also derived general kinetic equation~(\ref{ff1}) for the average
distribution function of energy losses in random fractal medium, that can be
considered as generalization of the Landau equation~(\ref{bk0}) for
homogeneous medium with constant density of scattering centers, $g\left(
r\right)  =1/a$. In heterogeneous medium the integral kernel $g\left(
r\right)  $ of this equation is proportional to the structure function
$G\left(  r\right)  $ of the medium. The non-local character of the kinetic
equation~(\ref{ff1}) is related to the presence of strong non-local
correlations in a fractal medium. The consideration of scattering of paricles
in turbulent medium needs additional study because of multifractal structure
of turbulent flows.

\appendix{}

\section{Estimation of exponent $\alpha$ for fractal medium\label{EXPONENT}}

\begin{figure}
[tbh]
\begin{center}
\includegraphics[
height=1.506in,
width=1.5475in
]%
{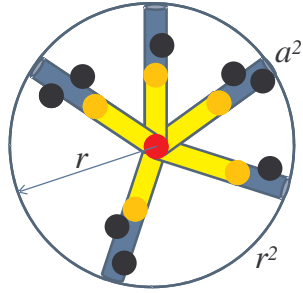}%
\caption{Scaling estimation of survival probability $\Psi\left(  r\right)  $.
The number of scattering centers inside the sphere of radius $r$ is $N\left(
r\right)  \simeq\left(  r/a\right)  ^{D}$. The particle collides first time
with only $N\left(  r\right)  \Psi\left(  r\right)  $ centers. The total
surface area of these centers $a^{2}N\left(  r\right)  \Psi\left(  r\right)
\simeq r^{2}$ equal to the area of the sphere. }%
\label{Psi1}%
\end{center}
\end{figure}

Here we present simple scaling estimation of the survival probability
$\Psi\left(  r\right)  $, Eq.~(\ref{Psi}), in the case of scattering of
particle in random medium with fractal dimension $D$. We consider all particle
trajectories colliding with the same scattering center as identical. Average
number of such trajectories starting from one scattering center can be found
by draw the sphere of radius $r$ around this center, see Fig.~\ref{Psi1}. The
total number of scattering centers inside this sphere is $N\left(  r\right)
\sim\left(  r/a\right)  ^{D}$. Note, that scaling consideration can be applied
to fractal objects with some care: to get the right scaling on the scale $r$
we have to place all these scattering centers randomly at a distance $\sim r$
from the center of the sphere. The propagating particle collides (first time,
as in the case of the first passage problem, see Fig.~\ref{Psi1}) only with
small survival part $\Psi\left(  r\right)  $ of these centers. Therefore, the
total number of different ballistic trajectories is $N\left(  r\right)
\Psi\left(  r\right)  $ and their total area on the sphere is $a^{2}N\left(
r\right)  \Psi\left(  r\right)  $. On the other hand, projections of these
trajectories on the sphere cover the whole area $\sim r^{2}$ of this sphere:%
\[
a^{2}N\left(  r\right)  \Psi\left(  r\right)  \simeq r^{2}%
\]
Solving this equation with respect to the survival probability we find%
\begin{equation}
\Psi\left(  r\right)  \simeq\frac{r^{2}}{a^{2}N\left(  r\right)  }%
\simeq\left(  \frac{a}{r}\right)  ^{D-2},\quad r>a.\label{Surv}%
\end{equation}%

\section{Wrize type functions\label{WRIZE}}

The Wrize function is defined as%
\begin{equation}
W_{\alpha}\left(  z\right)  =\sum_{l=0}^{\infty}\frac{\left(  -z\right)  ^{l}%
}{l!\Gamma\left(  1-\alpha-\alpha l\right)  } \label{Wa}%
\end{equation}
We can present this function in integral form using corresponding presentation
of Gamma function for non-integer $-z$%
\begin{equation}
\frac{1}{\Gamma\left(  z\right)  }=\frac{i}{2\pi}\int_{C}\left(  -t\right)
^{-z}e^{-t}dt \label{Gz}%
\end{equation}
The contour $C$ of integration encircles positive axis of complex variable
$t$. Substituting this expression into Eq.~(\ref{Wa}) and calculating the sum
over $l$ we find%

\begin{equation}
W_{\alpha}\left(  z\right)  =\frac{i}{2\pi}\int_{C}\left(  -t\right)
^{\alpha-1}e^{-t-z\left(  -t\right)  ^{\alpha}}dt \label{Wai}%
\end{equation}
The function $W_{\alpha}\left(  z\right)  $ is normalized by the condition%
\begin{equation}
\int_{0}^{\infty}W_{\alpha}\left(  z\right)  dz=\frac{i}{2\pi}\int_{C}\left(
-t\right)  ^{-1}e^{-t}dt=1
\end{equation}
Calculating the integral~(\ref{Wai}) by the steepest descent method, we find
its asymptotic behavior at large $z\gg1$:%
\begin{equation}
W_{\alpha}\left(  z\right)  \simeq\frac{1}{\sqrt{\pi\left(  1-\alpha\right)
}}\left(  z\alpha\right)  ^{-\frac{1-2\alpha}{2\left(  1-\alpha\right)  }%
}e^{-\frac{1-\alpha}{\alpha}\left(  z\alpha\right)  ^{\frac{1}{1-\alpha}}}%
\end{equation}

Explicit expressions can be obtained for some particular cases:%
\begin{equation}
W_{\alpha}\left(  z\right)  =\left\{
\begin{array}
[c]{ccl}%
e^{-z} & \text{for} & \alpha=0\\
\frac{1}{\sqrt{\pi}}e^{-z^{2}/4} & \text{for} & \alpha=1/2\\
\delta\left(  z-1\right)  & \text{for} & \alpha=1
\end{array}
\right.
\end{equation}

The one-parameter Mittag-Leffler function is defined by the series%
\begin{equation}
E_{\alpha}\left(  z\right)  =\sum_{l=0}^{\infty}\frac{\left(  -z\right)  ^{l}%
}{\Gamma\left(  1+l\alpha\right)  }. \label{Ea}%
\end{equation}
Its integral presentation can be found similar to Eq.~(\ref{Wai}):%
\begin{equation}
E_{\alpha}\left(  z\right)  =\frac{i}{2\pi}\int_{C}\frac{\left(  -t\right)
^{\alpha-1}}{z+\left(  -t\right)  ^{\alpha}}e^{-t}dt \label{Eaz}%
\end{equation}
and at large $z\gg1$ and $\alpha<1$ it decays as:%
\begin{equation}
E_{\alpha}\left(  z\right)  \simeq\frac{1}{\Gamma\left(  1-\alpha\right)  z}%
\end{equation}
We show also known explicit expressions for this function:%
\begin{equation}
E_{\alpha}\left(  z\right)  =\left\{
\begin{array}
[c]{ccl}%
\left(  1+z\right)  ^{-1} & \text{for} & \alpha=0\\
e^{z^{2}}\operatorname{erfc}\left(  z\right)  & \text{for} & \alpha=1/2\\
e^{-z} & \text{for} & \alpha=1
\end{array}
\right.
\end{equation}

\section{Solution of CTRW\label{ASYMPT}}

The solution of Eq.~(\ref{ff0}) with function $\tilde{g}\left(  q\right)
$~(\ref{gq}) has the form:%
\begin{equation}
\tilde{f}\left(  p,q\right)  =\frac{1}{q+\omega\left(  p\right)  \left(
aq\right)  ^{1-\alpha}} \label{phi}%
\end{equation}
where $\omega\left(  p\right)  =\left[  1-\tilde{w}\left(  p\right)  \right]
/a$. Calculating the inverse Laplace transform over $q$%
\[
\tilde{f}\left(  p,x\right)  \equiv\int_{0}^{\infty}d\Delta e^{-p\Delta
}f\left(  \Delta,x\right)
\]
we find in the limits of small and large $x$:

a) At small $x$ one can expand Eq.~(\ref{phi}) in powers of small $q^{-\alpha
}$, and we get simple analytical form of the Laplace transform of the
distribution function
\begin{equation}
\tilde{f}\left(  p,x\right)  =E_{\alpha}\left[  \omega\left(  p\right)
a^{1-\alpha}x^{\alpha}\right]  \label{fEa}%
\end{equation}
where $E_{\alpha}$ is the one-parameter Mittag-Leffler function, defined in
Eq.~(\ref{Eaz}) of Appendix~\ref{WRIZE}.

b) At large $x$ one can expand Eq.~(\ref{phi}) in powers of small $q^{\alpha}%
$, and we find%
\begin{equation}
\tilde{f}\left(  p,x\right)  =\sum_{l=1}^{\infty}\frac{\left(  -1\right)
^{l-1}}{\Gamma\left(  1-\alpha l\right)  }\frac{1}{\left[  a\omega\left(
p\right)  \left(  x/a\right)  ^{\alpha}\right]  ^{l}} \label{fS}%
\end{equation}

In the case of constant losses of particle energy $\Delta=\Delta_{1}n$ the
function $\omega\left(  p\right)  =\left[  1-e^{-p\Delta_{1}}\right]  /a$, and
we find from Eq.~(\ref{phi})%
\begin{equation}
\tilde{f}_{n}\left(  q\right)  =\frac{\left(  aq\right)  ^{\left(
1-\alpha\right)  n}}{\left[  q+\left(  aq\right)  ^{1-\alpha}\right]  ^{n+1}}
\label{fnq}%
\end{equation}

\end{document}